\documentclass[nofootinbib,preprint,
eqsecnum,
amsmath,amssymb,
amsmath,amssymb, aps,
]{revtex4}
\usepackage[switch]{lineno}

\usepackage{lineno} 
\usepackage[dvipsnames]{xcolor} 
\usepackage{epsfig}
\usepackage{amsfonts}
\usepackage{amsmath}
\usepackage{wasysym}
\usepackage{amssymb}
\usepackage{bm} 
\usepackage[framemethod=PStricks]{mdframed}
\usepackage{lipsum}
\usepackage{float}
\usepackage{soul}

\usepackage{pdflscape}
\usepackage{graphicx}
\usepackage{ulem}

\newcommand{\be}{\begin{equation}}
\newcommand{\ee}{\end{equation}} 
\newcommand{\bea}{\begin{eqnarray}} 
\newcommand{\eea}{\end{eqnarray}}

\usepackage[utf8]{inputenc}
\usepackage{bbm} 				% identity operator
\usepackage{epstopdf}				% self explanatory
\usepackage{verbatim}    			% long comments
\usepackage{sidecap}                % captions on the side
\usepackage{indentfirst}
\usepackage[colorinlistoftodos]{todonotes}

\definecolor{lightred}{rgb}{1.0, 0.13, 0.32}

\usepackage[utf8]{inputenc}
\usepackage[T1]{fontenc}
\usepackage{color}
\usepackage{float}
\usepackage{soul}
\usepackage[dvipsnames]{xcolor}
\usepackage{braket}
\newcommand\bes{\begin{equation*}}
\newcommand\ees{\end{equation*}}

\newcommand\refp[1]{({\ref{#1}})}

\definecolor{armygreen}{rgb}{0.0, 0.5, 0.0}

\newcommand{\vbib}[5]{{#1}, {#2} \textbf{{#3}}, {#4} ({#5}).}

\usepackage{amssymb}
\usepackage{amsmath}
\usepackage{amsthm}
\usepackage{hyperref}

\begin{document}

\title{Eddy-Viscous Modeling and the Topology of Extreme \\ Circulation Events in Three-Dimensional Turbulence}

\author{G.B. Apolin\'ario$^1$, L. Moriconi$^2$, R.M. Pereira$^3$, and
V.J. Valadão$^2${\footnote{Corresponding author: valadao@pos.if.ufrj.br}}}
\affiliation{$^1$ENS de Lyon, CNRS, Laboratoire de physique, F-69342 Lyon, France}
\affiliation{$^2$Instituto de F\'\i sica, Universidade Federal do Rio de Janeiro,
C.P. 68528, CEP: 21941-972, Rio de Janeiro, RJ, Brazil,}
\affiliation{$^3$Instituto de Física, Universidade Federal Fluminense, 24210-346 Niterói, RJ, Brazil.}
%\date{May 2020}
\vspace{1.5cm}

\begin{abstract}
We discuss the role of particular velocity field configurations -- instantons, for short -- which are supposed to dominate the flow during the occurrence of extreme turbulent circulation events. Instanton equations, devised for the stochastic hydrodynamic setup of homogeneous and isotropic turbulence, are applied to the interpretation of direct numerical simulation results. We are able in this way to model the time evolution of extreme circulation events for a broad range of scales, through the combined use of eddy viscosity phenomenology and exact creeping instantons. While this approach works well for the core of circulation instantons, it fails to describe their tails. In order to overcome this difficulty, we put forward a numerical treatment of the axisymmetric instanton equations. Circulation instantons are then found to have a surprising topological structure, which consists of a system of paired counter-rotating vortex rings centered around the symmetry axis of a background axisymmetric vortical flow.
\end{abstract}

\maketitle

%\date{}

\section{Introduction}\label{sec1}

Predicting extreme events is a fundamental problem in completely distinct domains of science and technology, as in astrophysics \cite{aschwanden2019self,larrodera2021estimation}, climatology \cite{bouchet2014stochastic,laurie2015computation}, wind power generation \cite{couto2021identification}, and a further broad list of other important contemporary fields such as finance \cite{rocco2014extreme}, meteorology \cite{yiou2004extreme,grotjahn2016north} and seismology \cite{soloviev2008transformation}. 
%such as finance \cite{rocco2014extreme}, astrophysics \cite{aschwanden2019self}, meteorology \cite{grotjahn2016north} and many others.
Extreme events often take place in relatively very short time scales, but their consequences
can be perennially disastrous as in the case of earthquakes, hurricanes and financial market crashes. 

Navier-Stokes dynamics, our specific focus in this work, is a particularly interesting instance for the study of extreme events.~The formation of intense energy-carrying localized structures in turbulent flows is a fingerprint of the non-local and nonlinear aspects of the underlying equations of motion. Fast spatial variations of the dissipation field or of the entangled system of vortex tubes which confine strong vorticity \cite{yoshida2000statistical,kaneda2003energy} are, however, difficult to observe, and decades went by until it was realized that the self-similar cascade picture of turbulence, originally proposed by Kolmogorov (K41) \cite{frisch}, should be revised to include the existence of such extreme events. As it is well known, the predicted K41 scaling exponents for the statistical moments of velocity differences fail to reproduce the respective numerically and experimentally measured values \cite{batchelor1949nature,anselmet1984high,vincent1991spatial} at high moment orders. Commonly understood as an important feature of the {\it{intermittency phenomenon}} \cite{frisch,kaneda2003energy}, these deviations of the K41 scaling exponents have been accurately obtained through multifractal formulations \cite{kolmogorov1962refinement,obukhov1962,novikov1994intermittency,she1994universal,apolinario2020vortex,pereira2022hard}, but more fundamental derivations are still in order.

%numerically \cite{balkovsky1997intermittency,chernykh2001large,moriconi2009instanton,grafke2015relevance,apolinario2019onset}, but observables in 3D turbulence has much less attempts on the application of different formalism on the way to understand intermittency \old{in standard three dimensional Navier-Stokes turbulence} \textcolor{blue}{(too repetitive??)}. 
A systematic analytical strategy to address the problem of intermittent velocity fluctuations was initially introduced in the framework of Burgers turbulence \cite{burgers1948mathematical}. Asymptotic results were found for the right tail of the velocity gradient probability distribution functions \cite{gurarie1996instantons}, based on the steepest-descent approximation as implemented in the Martin-Siggia-Rose-Janssen-de-Dominicis \cite{martin1973statistical,janssen1976lagrangean,dominicis1976technics} (MSRJD) functional formalism. Subsequent theoretical and numerical efforts have been made to describe the left tail of Burgers velocity gradient and velocity difference PDFs \cite{balkovsky1997intermittency,chernykh2001large,moriconi2009instanton,grafke2015relevance,apolinario2019onset}. Despite the relevant progress achieved for the Burgers and effective Lagragian turbulent models \cite{johnson2016closure,grigorio2017instantons,apolinario2019instantons,alqahtani2021extreme}, the description of intermittency in three-dimensional turbulence is still very open to investigation along the lines of the MSRJD approach.

In this work, we use the MSRJD functional formalism to study the scaling properties of velocity circulation in three-dimensional homogeneous and isotropic turbulence, motivated by the fact that velocity circulation fluctuations are known to provide clear signatures of turbulence intermittency \cite{umeki1993probability,migdal1994loop,cao1996properties,benzi1997self}. Velocity circulation is a natural tool to probe localized vortex structures, which contain most of the turbulent kinetic energy in high Reynolds number flows \cite{kaneda2003energy,frisch}. %Velocity circulation, which is an inviscid Lagrangian invariant [refs], is furthermore seen to play a central role in various fluid dynamics phenomena [refs]. 
The first formally structured attempt to explore velocity circulation as the main actor of turbulent fluctuations dates back to 1994 \cite{migdal1994loop}. Promising results about the circulation probability distribution functions (cPDFs) were then advanced, but further developments and validation studies faced serious obstacles in the following years, due to computational limitations. More recently, once hardware shortcomings were overcome, the subject of circulation statistics was vigorously revisited under the light of much improved high-resolution simulations, not only in classical \cite{iyer2019circulation,iyer2021area} but also in quantum turbulence \cite{muller2021intermittency,polanco2021vortex}. Interesting perspectives on the multifractal and structural views of classical turbulence cascades have been put forward since then \cite{apolinario2020vortex,pereira2022hard,moriconi2021multifractality,moriconi2022statistics}.

Throughout our considerations, the circulation variable is defined as
\be\label{circref}
\Gamma_R=\oint_{\mathcal{C}_R}  v_idx_i=\iint_{\mathcal{D_R}}\omega_idA_i \ , \
\ee
where $\mathcal{D_R}$ is a circular domain of radius $R$ and ${\mathcal{C}_R}$ is the circular contour that encloses $\mathcal{D_R}$. The fields $v_i$ and $\omega_i$ are, respectively, the velocity and vorticity fields related through $\omega_i=\epsilon_{ijk}\partial_jv_k$ (we use Einstein notation for the summation over repeated indices, unless explicitly stated). We are interested in discussing velocity field configurations -- instantons, for short -- which are associated to extreme circulation events, that is, large deviations of $\Gamma_R$. On one front of analysis, we model them as exact creeping flow structures \cite{creeping} related to turbulent eddy-viscosity \cite{boussinesq1877essai} time scales. The approach proves to be phenomenologically meaningful, once we validate it by means of extensive turbulence databases. On another front, we numerically solve the axisymmetric nonlinear instanton equations, to find that the circulation instantons have a remarkable topological structure described by the superposition of two counter-rotating vortex rings that share the same symmetry axis and carry opposite helicity in the presence of a background axial vorticity field. 
%As we will see, this solution has formal connections to the problem of quantum vortex ring propagation in superfluids.

This paper is organized as follows. Sec.~\ref{sec2} briefly reviews the main technical aspects of the MSRJD functional formalism. We, then, obtain closed analytical solutions for the circulation instantons in the viscous limit of the Navier-Stokes equations and show their somewhat unexpected usefulness to model the time evolution of extreme circulation events, from the analysis of direct numerical simulation (DNS) data. The nonlinear instanton equations are introduced in Sec.~III, and their numerical solutions are carefully discussed, with emphasis placed on the qualitative differences between them and the creeping solutions derived in Sec.~II. Finally, in Sec.~IV, we summarize our findings and indicate directions of further research.

\section{The functional approach and creeping instantons}\label{sec2}

% Many physical systems of relevance shares the large deviation principle \cite{dembo1998large}, from standard statistical mechanics \cite{touchette2009large,ferre2019large} to out of equilibrium systems such as finance \cite{avellaneda2003application} and turbulence \cite{fuchs2021instantons}, as a consequence, the dynamics leading to an extreme event follow a deterministic differential equation named ``instantons equations''. The solutions of those equations are often named ``instantons'' and they share the property of being the most relevant field configuration in the weak noise limit or far tails of the probability density function (PDF). Such fields can be thought of as generators of extreme events and their properties are closely related to the scaling law that dominates PDF tails.

\subsection{MSRJD Formalism}

Many physical systems which exhibit extreme events among their dynamical states are suitable for the application of large deviation techniques \cite{weiss1995introduction,evertsz1995large,ragone2018computation}. It is often possible to address in these cases the MSRJD methodology \cite{martin1973statistical,janssen1976lagrangean,dominicis1976technics}, where one typically deals with differential equations that lead to solutions -- the so-called instantons -- which dominate the probability distribution tails that describe the occurrence of extreme events. 

We are interested in applying these ideas to three-dimensional turbulence in its stochastic hydrodynamics formulation \cite{brenig1976stochastic}. The stochastic incompressible Navier-Stokes equations are written as
\be\label{SNSE}
\partial_tv_i+v_j\partial_jv_i-\nu\partial^2v_i=-\partial_iP+f_i \ , \
\ee
\be\label{SNSE2}
\partial_iv_i=0 \ , \
\ee
where $v_i=v_i(\vec{x},t)$ is the velocity field, $P=P(\vec{x},t)$ is the pressure field ensuring the incompressibility constraint (Eq.~\ref{SNSE2}), $\nu$ is the kinematic viscosity, and $f_i=f_i(\vec{x},t)$ is a zero-average Gaussian random field used to model large-scale forcing, correlated as
\be
\langle f_i(\vec{x},t)f_j(\vec{x}',t')\rangle=\delta_{ij}\chi(\vec{x}-\vec{x}')\delta(t-t') \ . \
\ee
The exact form of $\chi(\vec{x}-\vec{x}')$ is not phenomenologically relevant insofar as its correlation length $L$ is much larger than the investigated length scales. Assuming homogeneity and isotropy, the correlation kernel $\chi(\vec x - \vec x')$ depends only on $|\vec x - \vec x'|$ and is parametrized by its amplitude $D_0=\chi(\vec{0})$ and its correlation length $L$, which can be conveniently defined as $L^2=\chi(0)/\chi''(0)$.

In our particular application of the MSRJD formalism, we deal with a functional measure which is path-integrated over $v_i(\vec{x},t)$ and an auxiliary field $p_i(\vec{x},t)$ \cite{ivashkevich1997symmetries,grafke2015instanton}. Lagrange multipliers $Q(\vec{x},t)$ and $\lambda$ are also introduced, to ensure incompressibility and to fix the circulation $\Gamma_R$ at a given time instant, respectively. More concretely, the cPDF is evaluated in the MSRJD formalism as
\bea\label{cpdf}
&&\rho(\Gamma_R)=\left\langle\delta\left(\Gamma_R-\oint_{\mathcal{C}}v_i(\vec{x},0)dx_i\right)\right\rangle
= \nonumber \\
&&=\mathcal{N}^{-1}\int D[\vec{v}]D[P]D[\vec{p}]D[Q]\int_{-\infty}^{\infty}d\lambda\ \exp \left \{ -\frac{S[\vec{v},P,\vec{p},Q,\lambda]}{g^2} \right \} \ , \
\eea
where the pseudo-statistical weight (it is a complex number) contributions for $\rho(\Gamma_R)$ are given as the exponential of the MSRJD action, $S[\vec{v},P,\vec{p},Q,\lambda]$, specified below, and $\mathcal{N}$ is an unimportant normalization constant (to be suppressed hereafter). A single dimensionless parameter $g^2=\chi(\vec{0})L^4\nu^{-3}$ can be factored out from the action and its role is to control the noise strength and to set what should be considered an extreme event: these are the ones associated to MSRJD actions which are much larger than $g^2$. 

The above functional integration represents the probability density functional to find velocity circulation $\Gamma_R$ as defined in Eq.~(\ref{circref}), at a particular time instant, namely $t=0$, due to the cumulative effect of forcing since the remote past, say $t\rightarrow -\infty$. All fields are assumed to vanish in the remote past. Moreover, instantons are restricted to $t<0$, such that auxiliary fields are imposed to vanish at $t\rightarrow 0^+$ as a boundary condition \cite{chernykh2001large}. Measuring length in units of $L$ and time in units of the viscous time scale
\be
\tau_\nu = \frac{L^2}{\nu} \ , \
\ee
a dimensionless form of the MSRJD action can be written as
\begin{align}\label{actionre-scaled}
   && \tilde{S}[\vec{v},P,\vec{p},Q,\lambda] =-
    \frac{1}{2}\int d^3x d^3x' dt\  p_i(\vec{x},t) \tilde{\chi}(\vec{x}-\vec{x}') p_i(\vec{x}',t)+
    \int d^3x dt \ Q\partial_i v_i+\nonumber\\
    &&+\int d^3x dt \ p_i(\partial_tv_i+v_j\partial_jv_i+\partial_iP-\partial^2 v_i)
    -\lambda\left(\Gamma_{R/L}- \oint_{\mathcal{C}_{R/L}} \!\! dx_i v_i(\vec{x},t=0)\right) \ , \
\end{align}
where, now, $\tilde{\chi}(\vec{x}-\vec{x}')$ is the force-force correlator with unit correlation length. The connection between $g$ and the Reynolds number Re is made by the assumption that a well defined inertial range develops when the dissipation length $\eta_K=(\nu^3/\chi(\vec{0}))^{1/4}$ is much smaller than the correlation length $L$. Since the characteristic velocity scale is $U=(\chi(\vec{0})L)^{1/3}$, we obtain, as a consequence, that Re~$=UL/\nu=g^{2/3}$.

Instantons correspond to the extrema of the MSRJD action and are assumed to provide, from Eq.~(\ref{cpdf}), the dominant saddle-point contributions in the path-integral evaluation of extreme event probabilities. They are found as solutions of the variational principle $\delta \tilde{S}=0$, which amounts here to the instanton equations,
\be\label{instv}
(\partial_t-\partial^2)v_i+v_j\partial_j v_i+\partial_iP=
\int d^3x'\  p_i(\vec{x}',t) \tilde{\chi}(\vec{x}-\vec{x}') \ , \
\ee
\be\label{instp}
(\partial_t+\partial^2)p_i+ v_j\partial_j p_i+ v_j\partial_i p_j+\partial_iQ=\lambda\delta(t)\oint_{\mathcal{C}_{R/L}} \!\! dx'_i\delta^3(\vec{x}-\vec{x}') \ , \
\ee
subject to the constraints,
\be\label{instcv}
\partial_iv_i=0 \ , \
%\ee
%\be\label{instcp}
\partial_ip_i=0 \ , \
\ee
\be\label{instcirc}
\Gamma_{R/L}=\oint_{\mathcal{C}_{R/L}} \!\! dx'_iv_i(x',t=0) \ . \
\ee
By solving the above equations, one gets asymptotic expressions for the cPDF tails, from the mapping of the values of the Lagrange multiplier $\lambda$ to the final velocity circulation $\Gamma_{R/L}[\vec{v}(\lambda)]$. The instanton calculus yields, in this way,
\be
\rho(\Gamma_{R/L}) \simeq C \exp \left [ - \frac{1}{g^2} \tilde{S}_c(\lambda) \right ] \label{cPDF} %\ , \
\ee
as an asymptotic approximation that holds for $g \ll \tilde{S}_c(\lambda)$, where 
$\lambda = \lambda(\Gamma_{R/L})$ and $C$ is a normalization constant. A more mathematically rigorous discussion of the arguments leading to results which are analogous to (\ref{cPDF}) is the essential subject of large deviation theory \cite{dembo1998large}. As a direct consequence of the Gärtner-Ellis theorem, a probability density function necessarily follows the {\it{Large Deviation Principe}} (Eq.~(\ref{cPDF}), in our case) if the associated action is strictly convex and the map between $\lambda$ and the observable of interest is differentiable for an arbitrary $\lambda$. Non-convex actions or ill-defined $\lambda$-maps can be nevertheless adapted to large deviation evaluations through the use of alternative integration measures \cite{alqahtani2021instantons}.

\subsection{Creeping Instantons and Eddy Viscous Modeling}\label{S2.2}

Instanton equations are rarely amenable to analytical treatment without the help of phenomenological or empirical inputs about the underlying dynamics. Approximate solutions of (\ref{instv}-\ref{instcirc}) and of analogous magnetohydrodynamic equations  were formerly carried out under restrictive assumptions about the roles of symmetries and the strain rate field \cite{moriconi1998circulation,moriconi2002circulation}. Somewhat surprisingly, as we will discuss in the following, interesting information on extreme events can be obtained from the exact form of viscous (creeping) instantons.

\begin{figure}[t]
\centering\includegraphics[width=0.9\linewidth]{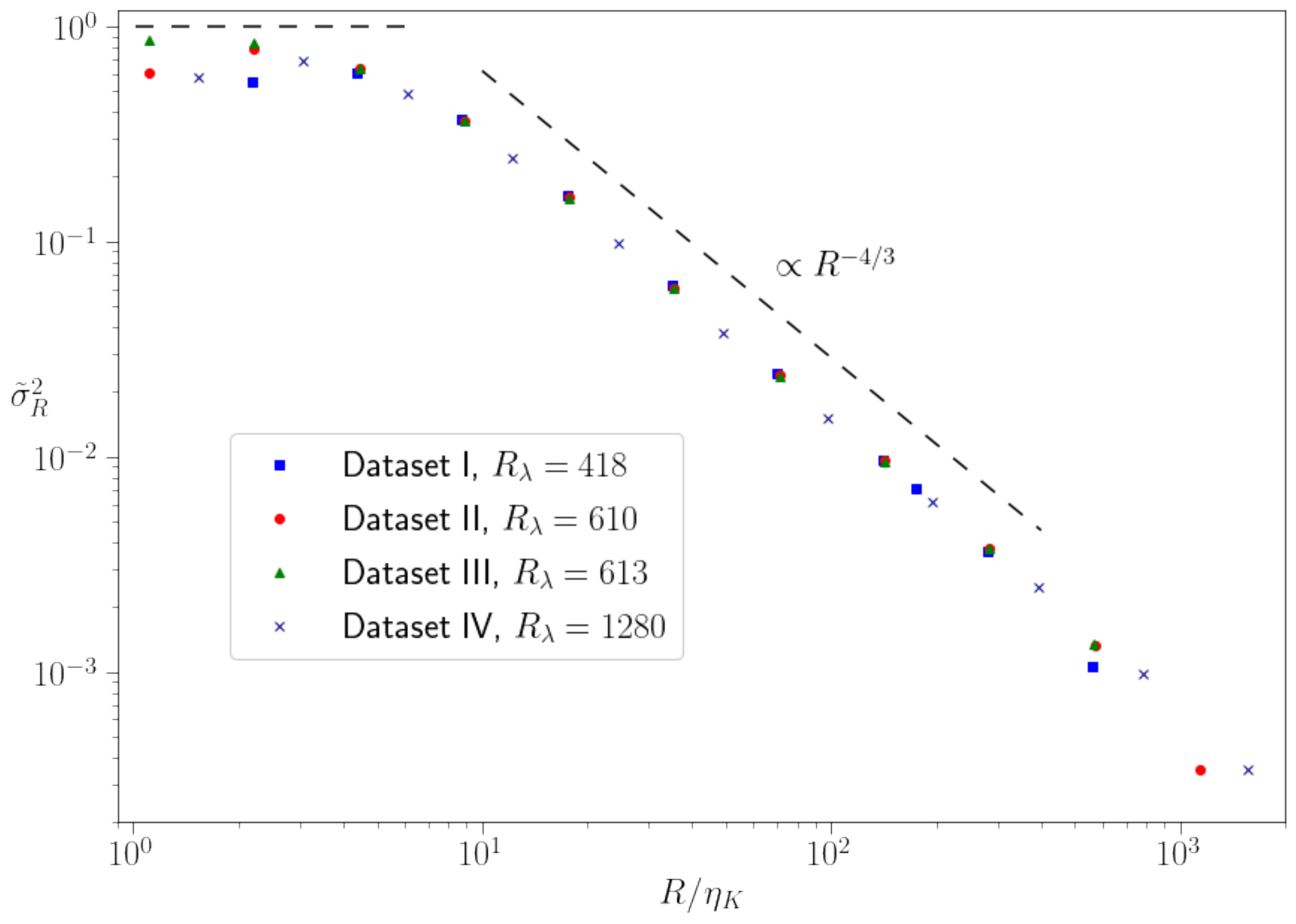}
\caption{The dimensionless variance $\tilde \sigma_R^2$, defined in (\ref{sigmaR}), as a function of the contour radius $R$ for various Reynolds numbers (symbols). Dashed lines represent $\tilde \sigma_R^2=1$ and $\tilde \sigma_R^2 \propto R^{-4/3}$.}
\label{fig1}
\end{figure}

To work out the creeping instantons, we introduce the diffusion Green's function with vanishing boundary condition in $\mathbb{R}^3$,
\be
G(\vec{x},t)=\frac{1}{(4\pi|t|)^{-\frac{3}{2}}}\exp \left (- \frac{|\vec{x}|^2}{4|t|} \right ) \ . \ 
\ee
Neglecting the nonlinear terms in Eqs.~(\ref{instv}-\ref{instcirc}), the velocity and the conjugate solutions read
\be\label{linv}
v^{(1)}_i(\vec{x},t)=\int_{-\infty}^tdt'\int d^3x'd^3x''
G(\vec{x}-\vec{x}',t-t') \tilde{\chi}(\vec{x}'-\vec{x}'')p_i^{(1)}(\vec{x}'',t') \ , \
\ee
\be\label{linp}
p^{(1)}_i(\vec{x},t)=\lambda\oint_{\mathcal{C}_{R/L}}dx_i''\int_{t}^0
dt'\int d^3x'
G(\vec{x}-\vec{x}',t-t')\delta(t')\delta^3(\vec{x}'-\vec{x}'') \ . \
\ee\label{linac}
The MSRJD action, Eq.~(\ref{actionre-scaled}), additionally reduces to
\be
S^{(1)}_0[v_i,p_i]=\frac{\lambda}{2g^2}\int_{\mathcal{C}_{R/L}}dx_iv_i^{(1)}(\vec{x},0)=\frac{\lambda \Gamma^{(1)}_{R/L}(\lambda)}{2g^2} \ . \
\ee
Defining, at this point, the explicit form of the force-force correlator as
\be
\tilde{\chi}(\vec{x}-\vec{x}') = \frac{1}{(2\pi)^{3/2}}\exp\left(-\frac{|\vec{x}-\vec{x}'|^2}{2}\right) \ , \
\ee
we obtain from (\ref{linv}) and (\ref{linp}) the creeping instanton solution by setting $x_3$ to be perpendicular to the contour $\mathcal{C}_{R/L}$
\be\label{velcomp}
v^{(1)}_i(\vec{x},t)=\lambda\pi\sqrt{\frac{R/L}{x_\perp}}\epsilon_{3ji}\frac{x_j}{x_\perp}
\int_0^{\sqrt{\frac{x_\perp R/L}{1-2t}}} du \ e^{-\left(\frac{(R/L)^2+x_\perp^2+x_3^2}{2x_\perp (R/L)}\right)u^2}I_1(u^2) \ , \
\ee
where $I_1(\cdot)$ is the modified Bessel function of the first kind and $x_\perp^2=x_1^2+x_2^2$. We remark that part of the spatio-temporal dependence of the velocity field is encoded in the upper limit of the above integration. It is interesting to note that this integral can be exactly computed only at points that belong to the contour ${\mathcal{C}_{R/L}}$, defined by $x_3 =0$ and $|x_\perp| = R/L$. We find, in cylindrical coordinates,
\be\label{velinstanton}
\vec{v}^{(1)}(\mathcal{C}_{R/L},t)=\lambda\frac{\pi}{6} \frac{R^3/L^3}{(1-2t)^{3/2}}\ {}_2F_2\left(\frac{3}{2},\frac{3}{2},\frac{5}{2},3;-2\frac{R^2/L^2}{1-2t} \right)\hat{\theta} \ , \
\ee
where ${}_2F_2$ is the hypergeometric function of $2+2$ entries. The time-dependent instanton circulation is readily obtained as
\be\label{timeinstanton}
\Gamma^{visc}_R(t) \equiv \oint_{\mathcal{C}_{R/L}}\vec{v}^{(1)}(\vec{x},t)\cdot d\vec{x}=\lambda\frac{\pi^2}{3} \frac{R^4/L^4}{(1-2t)^{3/2}}\ {}_2F_2\left(\frac{3}{2},\frac{3}{2},\frac{5}{2},3;-2\frac{R^2/L^2}{1-2t} \right) \ . \
\ee
Since the circulation is a linear function of $\lambda$, the resulting cPDF is Gaussian. Recovering the original circulation units, its variance $\sigma_R^2$ is found to satisfy
\be\label{variance}
\frac{\sigma_R^2}{\nu^2 {\hbox{Re}^3}}=\frac{\pi^2}{3} \frac{R^4}{L^4}\ {}_2F_2\left(\frac{3}{2},\frac{3}{2},\frac{5}{2},3;-2\frac{R^2}{L^2} \right) \ . \
\ee
%Eq.~(\ref{variance}) extends the previous result of \cite{moriconi2002circulation} where no specific form of the correlation is adopted, for an analysis of fluctuations in the dissipative region $R\leq \eta_K$. 
We have, for $R \ll L$, 
\be
{}_2F_2\left(\frac{3}{2},\frac{3}{2},\frac{5}{2},3;-2\frac{R^2}{L^2} \right) = 1 - \frac{3}{5} \left ( \frac{R}{L} \right )^2 + {\cal{O}}((R/L)^4) \ , \
\ee
such that, in this asymptotic limit, 
\be
\tilde \sigma_R^2 \equiv \frac{3 \sigma_R^2 L^4}{(\pi R^2)^2 \nu^2 {\hbox{Re}}^3} \simeq 1 \ . \  \label{sigmaR}
%\sigma_R^2 \propto \nu R^4 \ . \  \label{sigmaR}
\ee
Although (\ref{sigmaR}) is expected to hold only for creeping flows, it may be possibily relevant even for higher Reynolds number solutions,
%Inertial forces are suppressed in the vortex cores, which have typical linear sizes of a few Kolmogorov dissipation lengths [refs]. Therefore, 
once scales are probed and the MSRJD path-integration (\ref{cpdf}) is dominated by smooth velocity field configurations. The educated guess that generalizes (\ref{sigmaR}) to turbulent flows sustained by alternative forcing mechanisms, where $L$ and Re are not necessarily defined in terms of Gaussian force-force correlators, reads
\be
\frac{3 \sigma_R^2 L^4}{(\pi R^2)^2 \nu^2 {\hbox{Re}}^3} \simeq {\cal{O}}(1) \ . \  \label{sigmaRp}
%\sigma_R^2 \propto \nu R^4 \ . \  \label{sigmaR}
\ee
To investigate the correctness of (\ref{sigmaRp}), we have worked with four DNS datasets of homogeneous and isotropic turbulent flows, publicly available from the Johns Hopkins Turbulence Database (JHTDB) platform\footnote{For more information about the public datasets see {\url{http://turbulence.pha.jhu.edu/}}.} \cite{perlman2007data,JHTD2,JHTD3,JHTD4}. The simulations were developed in periodic cubic lattices with dimensions $1024^3$ ({\hbox{dataset I}}), $4096^3$ (datasets II), and $8192^3$ (datasets III and IV), corresponding to Taylor-based Reynolds numbers 418, 610, 613 and 1280 for datasets from I to IV. We have produced from these datasets ensembles which contain respectively $1\times 10^7$, $8 \times 10^6$, $9 \times 10^6$, and $4.5 \times 10^7$ circulation samples, computed by employing the first equality of Eq.~(\ref{circref}). Denoting by $\eta_K$ the Kolmogorov dissipation length \cite{frisch}, the results reported in Fig.~\ref{fig1} fully corroborate our expectations: relation (\ref{sigmaRp}) is in fact verified for $R < \eta_K$ (it should be noted that small scale evaluations of $\sigma_R$ are subject to relatively larger error bars, due to the inaccuracy associated to the representation of a circular contour in a square lattice, as closed polygonal line).

Fig.~\ref{fig1} also shows that the LHS of (\ref{sigmaRp}) leads, incidentally, to an excellent collapse of data across the inertial range scales $30 < R / \eta_K < 300$. Furthermore, the circulation variance has, in this range, a scaling dependence with $R$ which is very well approximated by the K41 scaling exponent $8/3$ \cite{iyer2019circulation}, as indicated by a dashed line Note that at fixed integral scale and fixed energy injection rate per unit mass, $\epsilon = U^3/L$, relation (\ref{sigmaRp}) implies that $\sigma^2_R \propto \nu^{-1} R^4$. Insisting that an analogous result should hold in the inertial range, we replace $\nu$ in this last relation by the scale-dependent eddy viscosity \cite{smith1998renormalization},
\be
\nu_R = \nu_0 \left ( \frac{R}{R_0} \right )^\frac{4}{3} \ , \ \label{eddyv}
\ee
where $\nu_0$ and $R_0$ are some reference viscosity and length scale parameters, to obtain the observed scaling $\sigma^2_R \propto \nu_R^{-1} R^4 \propto R^{8/3}$.

As a brief historical digression, we point out that the concept of eddy viscosity, long ago conceived by Boussinesq \cite{boussinesq1877essai} and much later revived through the distant interaction approximation (DIA) formalism \cite{kraichnan1964decay} and dynamical renormalization group techniques \cite{forster1977large,yakhot1986renormalization}, is essentially an effective transport parameter that allows for a coarse-grained description of the turbulent cascade. This is the main physical motivation for a number of numerically tractable models of turbulence, as the celebrated Smagorinsky sub-grid formulation of the Navier-Stokes equations \cite{smagorinsky1963general,scotti1993generalized}, which underlies the whole field of large eddy simulations \cite{germano1991dynamic,mason1994large,piomelli1999large,piomelli2014large}.

The replacement of the molecular viscosity $\nu$ by the eddy viscosity $\nu_R$ in (\ref{sigmaRp}) sounds at this point like a purely rhetoric remark. It is possible, however, that creeping instantons, when combined with eddy viscosity ideas, can in fact be used to model the time evolution of high Reynolds number circulation instantons. The argument goes as follows. Consider the set of all flow realizations which have evolved from the remote past and ended with extreme circulation 
\be\label{bins}
\Gamma_R \in [ \bar \Gamma_R(0) - \delta \Gamma, \bar \Gamma_R(0) + \delta \Gamma] \,\
\ee
at time $t=0$, where 
$\bar \Gamma_R(0) \gg \delta \Gamma > 0$ ($\delta \Gamma$ is just a measurement bin). Let now $\bar \Gamma_R(t)$ represent the time averaged circulation taken over the ensemble of all of these flow realizations, which we refer to as the {\it{filtered instanton circulation}}. 
Resorting to dynamic similarity and to the fact that $\eta_K/L \ll 1$ (i.e., Reynolds number is high), we write down the dimensionless circulation ratio (time units are restored to the original ones, from now on),
\be
\hat \Gamma_R(t) \equiv \frac{\bar \Gamma_R(t)}{\bar \Gamma_R(0)} = f(t/\tau_\nu, R/L, \eta_K / L) \simeq 
f(t/\tau_\nu, R/L, 0) \equiv \tilde f(t/\tau_\nu, R/L) \ . \ \label{ratiog}
\ee
It follows from Eq.~(\ref{ratiog}) that
\be
\hat \Gamma_R(\nu_R t / \nu) \simeq \tilde f(t/\tau_{\nu_R}, R/L) \ . \ \label{ratiog2}
\ee
In consonance with eddy viscosity phenomenology, we assume that Eq.~(\ref{ratiog2}) is scale invariant, viz.,
\be
\frac{d}{dR} \tilde f(t/\tau_{\nu_R}, R/L) =  0 \ . \ \label{rgeq}
\ee
We note that Eq.~(\ref{rgeq}) is just a renormalization group equation \cite{smith1998renormalization}, which has, as general solution, the functional relationship
\be
\tilde f(t/\tau_{\nu_R}, R/L) \equiv h(t/\tau_\nu) \ . \ \label{f=h}
\ee
This leads us, from (\ref{ratiog}), to
\be
\hat \Gamma_R(t) \simeq
\tilde f(t/\tau_\nu, R/L) = h(t'/\tau_\nu) \ , \ \label{f-hp}
\ee
with
\be
t' = t \frac{\nu}{\nu_R} = t \frac{\nu}{\nu_0} R^{-\frac{4}{3}} \ , \ \label{tp}
\ee
where we have used (\ref{eddyv}) to make the $R$-dependence explicit in the definition of $t'$.

In an analogous fashion, one has for the exact viscous solution (\ref{timeinstanton})
\be
\hat \Gamma^{visc}_R(t) \equiv  \frac{\Gamma^{visc}_R(t)}{\Gamma^{visc}_R(0)} = g(t/\tau_\nu, R/L) \equiv 
\frac{1}{(1-2t/\tau_\nu)^{3/2}}
\frac{{}_2F_2\left(\frac{3}{2},\frac{3}{2},\frac{5}{2},3;-2\frac{R^2/L^2}{1-2t/\tau_\nu} \right)}{{}_2F_2\left(\frac{3}{2},\frac{3}{2},\frac{5}{2},3;-2\frac{R^2}{L^2} \right)} 
\ . \
\ee
It is natural to conjecture, furthermore, that in the limit $R \rightarrow \eta_K$, the filtered and the viscous instanton circulations match, that is,
\bea
&&\hat \Gamma_{\eta_K}(t) =
\tilde f(t/\tau_{\nu}, \eta_k/L)
= 
\hat \Gamma^{visc}_{\eta_K}(t)
= g(t/\tau_\nu, \eta_K/L) \nonumber \\
&&\simeq g(t/\tau_\nu, 0) \equiv \tilde g(t/\tau_\nu) =\frac{1}{(1-2t/\tau_\nu)^{3/2}}
\frac{{}_2F_2\left(\frac{3}{2},\frac{3}{2},\frac{5}{2},3;0 
\right )}{{}_2F_2\left(\frac{3}{2},\frac{3}{2},\frac{5}{2},3;0 \right)} 
\ . \ \label{ratiog3}
\eea
Pushing the validity of Eq.~(\ref{f=h}) down to dissipative scales,  we get
\be
\tilde f(t/\tau_{\nu_R}, R/L) =  \tilde f(t/\tau_{\nu}, \eta_k/L) \ , \
\ee
which, in view of (\ref{ratiog3}), leads to
\be
\tilde f(t/\tau_{\nu_R}, R/L) \simeq   \tilde g(t/\tau_\nu) \ . \ \label{f=g}
\ee
Comparing (\ref{f=g}) with (\ref{f=h}), it follows that
\be
h(t/\tau_\nu) \simeq \tilde g(t/\tau_\nu) \ , \ 
\ee
so that $\hat \Gamma_R(t)$ can be modeled, from (\ref{f-hp}), as the viscous instanton solution $\tilde g(t'/\tau_\nu)$, that is,
\be
\hat \Gamma_R(t) = \hat \Gamma^{visc}_{0}(t') =
\frac{1}{(1-2t'/\tau_\nu)^{3/2}}
\frac{{}_2F_2\left(\frac{3}{2},\frac{3}{2},\frac{5}{2},3;0 
\right )}{{}_2F_2\left(\frac{3}{2},\frac{3}{2},\frac{5}{2},3;0 \right)} \ . \ \label{h-g}
\ee
In order to perform validation tests of Eqs.~(\ref{f-hp}) and (\ref{h-g}), we have implemented a numerical filtering procedure to extract, from the raw data of the JHTDB, the past time evolutions of extreme circulation events, which we detail below.

\subsection{Numerically Filtered Circulation Instantons}\label{S2.2}

We have taken Dataset I for statistical analyses, since this is the dataset which allows us to work with the largest ensemble of circulation time series, defined for circular contours centered at equally spaced grid points. Letting $\ell$ be the lattice parameter ($\ell\approx 2.1 \eta_K$), we have studied circulation fluctuations for four different radii, namely, $R= 8\ell$, $16\ell$, $32\ell$, and $64\ell$. For each given radius, we considered $32^3$ circular contours oriented normally to each Cartesian direction, for a total of $3 \times 32^3$ contours per radius.

We conventionally define the set $\Lambda_n$ of extreme events as the ensemble of circulations events whose absolute values $|\Gamma_R|$ reach some multiple $n$ of the circulation standard deviation, $\sigma_R=\sqrt{\langle \Gamma_R^2 \rangle}$, within a small tolerance window. In the language of Eq.~\refp{bins}, we consider the interval defined by $\bar\Gamma_R(0) = n \sigma_R$ with a 0,5\% tolerance, that is, {\hbox{$\delta\Gamma = 5 \times 10^{-3} \bar\Gamma_R(0)$}}.

\begin{figure}[h]
\centering\includegraphics[width=0.9\linewidth]{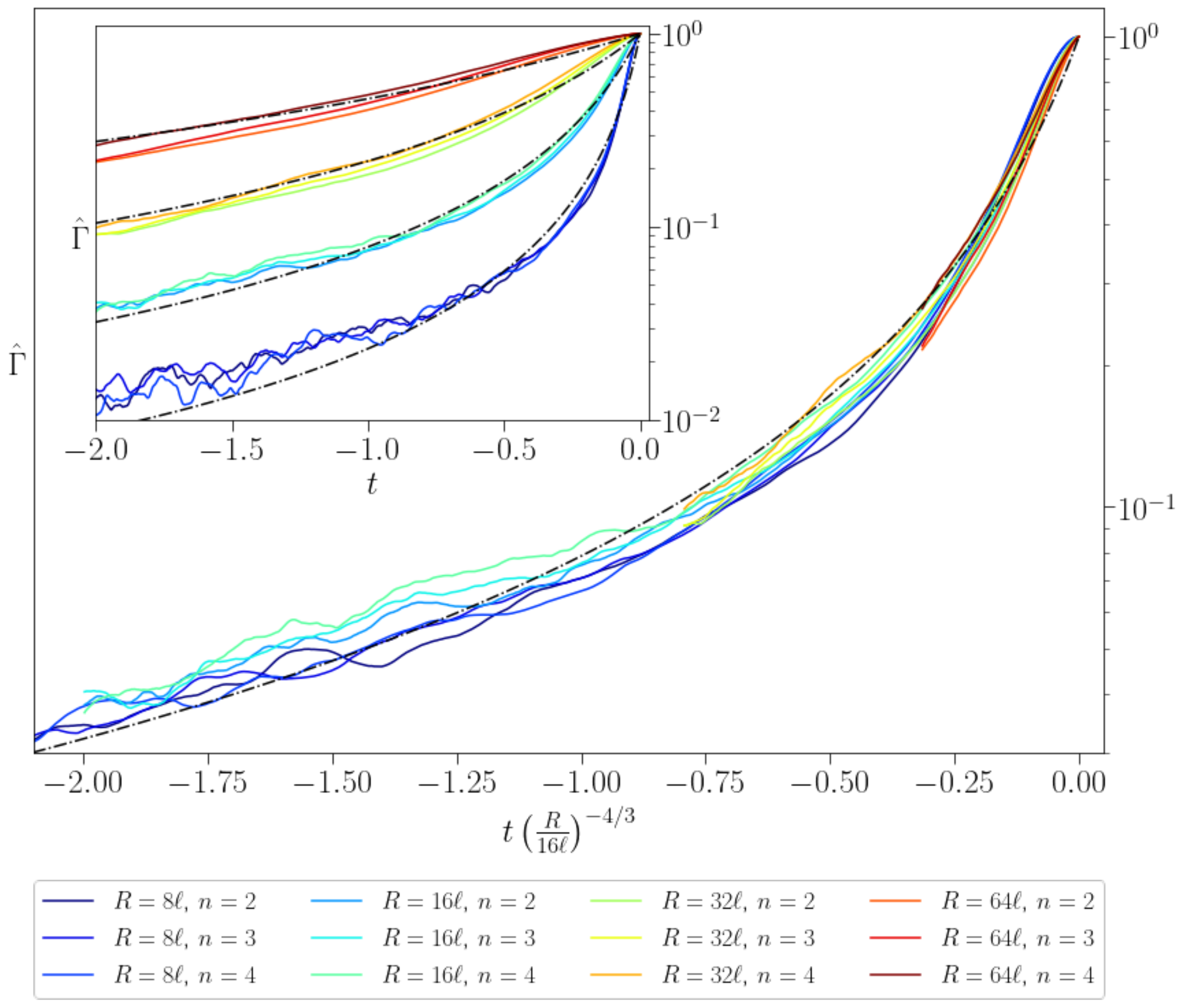}
\caption{Filtered circulation instanton $\hat \Gamma_R(t) \equiv \bar \Gamma_R(t) / \bar \Gamma_R(0)$ as a function of the rescaled time $t' = t (R/16 \ell)^{-4/3}$ for a number of radii and circulation sets $\Lambda_n$ at $R_\lambda = 433$ (solid lines). The eddy viscous modeling of the collapsed data is attained with the help of Eq.~(\ref{h-g}) (dash-dotted line). Inset: the corresponding non-collapsed plots, as functions of the original time variable $t$.}
\label{fig2}
\end{figure}

Once an extreme circulation event belonging to a given set $\Lambda_n$ is identified, we assign it the observation time instant $t=0$ and save its earlier time evolution. Then, a time dependent average $\bar \Gamma_R(t)$ over all saved series for each set $\Lambda_n$ is computed. Similar filtering procedures have been applied in instanton studies of Burgers turbulence \cite{grafke2013instanton}, Lagrangian turbulence models \cite{grigorio2017instantons,apolinario2019instantons} and rogue wave formation \cite{dematteis2018rogue,dematteis2019experimental}. 

Our results for $\hat \Gamma_R(t)$ are shown in Fig.~\ref{fig2}. Taking $R_0 = 16 \ell$ as an arbitrary reference length scale, we find that plots of $\hat \Gamma_R(t)$ as a function $t(R/16 \ell)^{-4/3}$ collapse reasonably well for all the investigated radii and sets $\Lambda_n$, as predicted by (\ref{f-hp}). Following now Eq.~(\ref{h-g}), we carry out an $L^2$-norm minimization of
\be
\sum_R ||\hat \Gamma_R(\nu_R t / \nu) - \hat \Gamma^{visc}_0(t) || \ , \ 
\ee
to adjust $\nu_0 = 4.5 \times 10^{-5} \nu$ as the reference viscosity in (\ref{eddyv}). Meaningful comparisons are then also reported in the same Fig.~\ref{fig2} for the eddy viscous modeling of $\hat \Gamma_R(t)$, as suggested by (\ref{tp}) and (\ref{h-g}). 

\begin{figure}[t]
\centering\includegraphics[width=0.9\linewidth]{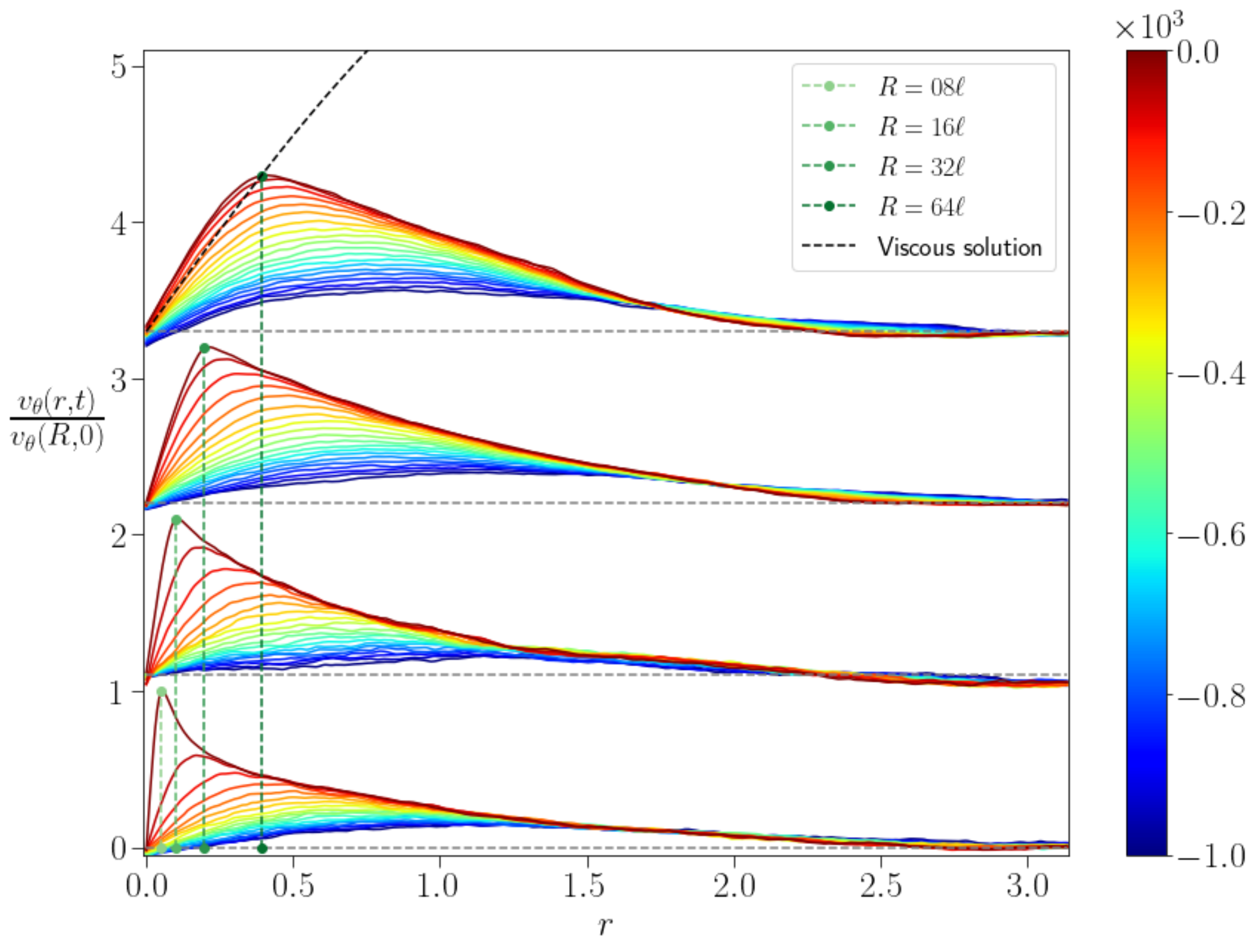}
\caption{Normalized filtered azimuthal velocity $\bar v_\theta(r,0,t)$ obtained from DNS data
for different radii $R$ (vertically shifted for clarity). Time evolution is color-coded (in units of the simulation time step) and vertical lines indicate $r=R$ in each simulation. For comparison purposes, part of a 
viscous instanton solution (dashed curve) is displayed for 
$R=64 \ell$.}\label{fig3}
\end{figure}

In addition to the eddy viscous modeling of the filtered circulations $\hat \Gamma_R(t)$ discussed above, one may wonder whether the functional forms of {\it{filtered instanton velocity fields}} could be effectively modeled by viscous instantons as well. With this aim, we take the symmetry axis of each prescribed circular contour as the $z$-axis of a local Cartesian reference system, and orient it in such a way that velocity circulations are rendered positive at the event occurrence time. Then, we perform a similar time-dependent averaging procedure over all events in $\Lambda_n$, as we did for $\hat\Gamma_R(t)$, but now for velocity field configurations in cylindrical coordinates, $\vec v = v_r \hat r + v_\theta \hat \theta + v_z \hat z$.
Due to isotropy, the mean velocity components are functions of $r$, $z$, and $t$ only, i.e., $\langle v_r \rangle \equiv \bar v_r(r,z,t)$, $\langle v_\theta \rangle \equiv \bar v_\theta(r,z,t)$, and $\langle v_z \rangle \equiv \bar v_z(r,z,t)$. 

Unfortunately, the radial and axial components $\bar v_r(r,z,t)$ and $\bar v_z(r,z,t)$ turn out to develop relatively small intensities, which prevents us from extracting a clear behavior out of the noise. The mean azimuthal component $\bar v_\theta(r,0,t)$, on the other hand, was found to be well-resolved in all studied cases. Fig.~\ref{fig3} shows comparisons between the numerically filtered $\bar v_\theta(r,t)$ (normalized by $\bar v_\theta(R,0)$) and the analogous normalized azimuthal velocity obtained from the viscous solution (\ref{velcomp}), where the dimensionless time $t/\tau_\nu$ is substituted by $t' / \tau_\nu$, as defined in Eq.~(\ref{tp}). It is seen that the eddy viscous modeling only provides an adequate description of $\bar v_\theta(r,t)$ for $r \leq R$, which is just the core region of the filtered instantons. Notwithstanding such a limitation, extreme circulations events are in fact well accounted by eddy viscous modeling, since, as the filtering results indicate, $\bar v_\theta(r,0)$ gets its maximum value at $r=R$.

%Now, regarding topological properties of the circulation instanton it is natural to expect that as $t\rightarrow 0$ the velocity field approaches its maximum value exactly at the contour $\mathcal{C}_{R/L}$ due to the narrow forcing term in Eq.~\ref{instv}. This is verified in Figure \ref{fig5} for the filtered instanton, but not for the viscous, one can easily show using Eq.~\ref{velinstanton} that $r=R$ is not a maximum of $v_\theta$.

%The conclusion drawn for renormalized viscous circulation instanton was that it describes low order statistics for a broad range of tested Reynolds numbers considering the picture of eddy viscosity for scales bigger than $4\eta_K$. Moreover, concerning the dynamics, it can model remote past properties of the full instanton when the later is time-rescaled, failing to reproduce spatial properties of such events. 

To further investigate the detailed structure of the circulation instantons moving beyond the viscous solutions, one must deal with the full nonlinear Euler-Lagrange equations (\ref{instp}-\ref{instcirc}), an issue we address in the next section. 

%In particular, the role of the scales $R$ and $L$ might be enlightening for the classification of intermittent fluctuations. To do so, we introduce a numerical method to solve the nonlinear instanton equations with appropriate boundary conditions. 

\section{nonlinear Instantons}\label{sec3}
% \textcolor{Fuchsia}{
% - Validacao das soluções analíticas de instanton viscoso. a fazer
% }

% \textcolor{Fuchsia}{
% - Soluções não-lineares: estrutura topológica, conexão com a estrutura tripla de anéis de vorticidade em superfluidez. a fazer
% }

A numerical scheme to solve the hydrodynamic instanton equations, similar to Eqs.~(\ref{instv}-\ref{instp}), was introduced by Chernykh and Stepanov \cite{chernykh2001large} in the context of Burgers turbulence, being later extended to a variety of models \cite{dematteis2018rogue,dematteis2019experimental}. In our case, the solution algorithm inserts, at ``iteration step $n$'', a given velocity field 
in Eq.~(\ref{instp}) to solve it backward in time. Its solution is defined as the auxiliary field $\vec p(\vec r,t)$ at iteration step $n$. This, in turn, is used in Eq.~(\ref{instv}) to provide the velocity field at iteration step $n+1$. An iteration cycle proceed recursively, until a prescribed convergence for the instanton fields is attained. The numerical procedure may start off with a vanishing velocity field or one may alternatively use some improved guess, as we do by using the creeping instanton described by (\ref{velcomp}).

We have worked out solutions for the circulation instantons in cylindrical coordinates with
Chebyshev collocation points (see the appendix for details) for a broad range of $\lambda$ values and contour radii. The numerical solutions were obtained through the Chernykh-Stepanov method outlined above, enhanced, at larger values of $\lambda$, by the convexification procedure discussed in \cite{alqahtani2021instantons}, in order to avoid possible spoiling inflection effects in the derivation of the cPDF tails. This amounts, in practical terms, to perform the replacement of $\Gamma$ by the {\it{nonlinear tilted measure}} ${\hbox{sign}}(\Gamma) \ln |\Gamma|$ in the MSRJD action (\ref{actionre-scaled}).

\begin{figure}[b]
\centering\includegraphics[width=0.9\linewidth]{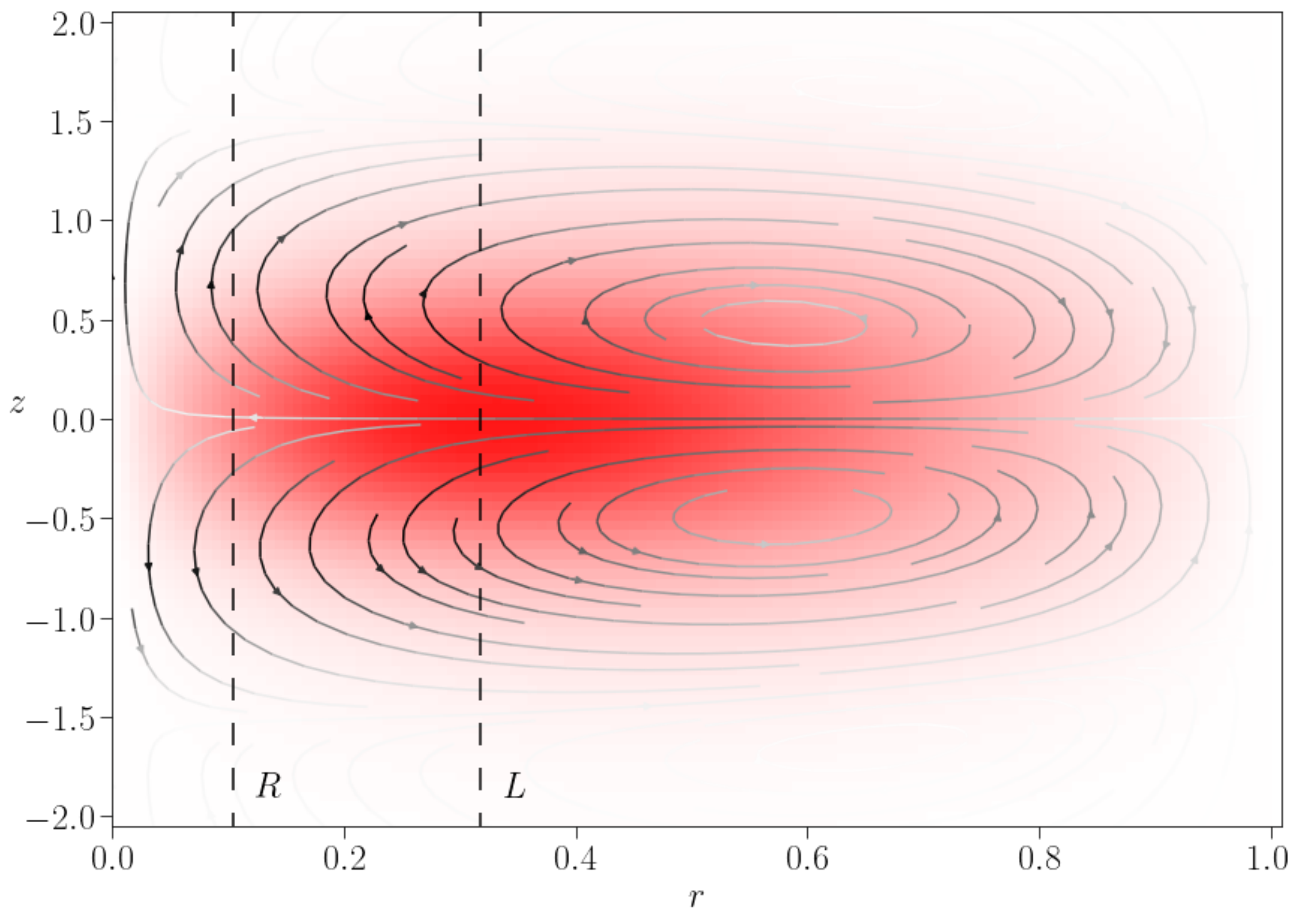}
\caption{Vector streamlines of a circulation instanton at its final evolution time on the $(r,z)$ plane. Values of $\sqrt{v_r^2(r,z) + v^2_z(r,z)}$ are qualitatively represented by the gray level of the streamlines, while the intensity of the azimuthal velocity $v_\theta(r,z)$ by a (red-colored) density plot. Values vary from 0 (white) to approximately $11.9$ (black) and $79.3$ (red), in units of $\nu/L$. This flow configuration is obtained as a solution of Eqs.~(\ref{instv}-\ref{instcirc}) for $\lambda=3\times 10^6$ and $R/L\approx0.32$. 
The vertical dashed lines give the radial positions $r=R$ and $r=L$.}
%Velocity field profile at $t=0$ as functions of $r$ and $z$ for $N=128$, $M=256$, $N_t=400$, $dt=1.25\times 10^{-3}$, $R/L\approx0.32$ and $\lambda=3\times 10^6$. Streamlines of $(v_r,v_z)$ are colored by intensity from zero (white) to its maximum value (black), red shaded area represents the field $v_\theta$. {\textcolor{blue}{Revisar}}
\label{fig4}
\end{figure}

Instability issues are known to affect the convergence of the Chernykh-Stepanov solutions at large values of the Lagrange multiplier $\lambda$ \cite{alqahtani2021instantons}. Although a number of technical improvements have been implemented to solve instanton equations in more efficient ways \cite{grafke2014arclength,grafke2019numerical,grigorio2020parametric}, three-dimensional instanton equations are far more computationally expensive than their dimensionally reduced counterparts.

Fig.~\ref{fig4} yields the typical velocity profile of a circulation instanton. A threefold vortex structure emerges, with interesting topological properties. It consists of two paired counter-rotating vortex rings, as they can be clearly identified from the streamline portrait there depicted in the $(r,z)$ plane. These rings define regions of opposite helicities, a fact associated to the background axisymmetric velocity field that circulates around the symmetry axis of the flow.
\vspace{0.2cm}

\begin{figure}[h]
\centering\includegraphics[width=0.9\linewidth]{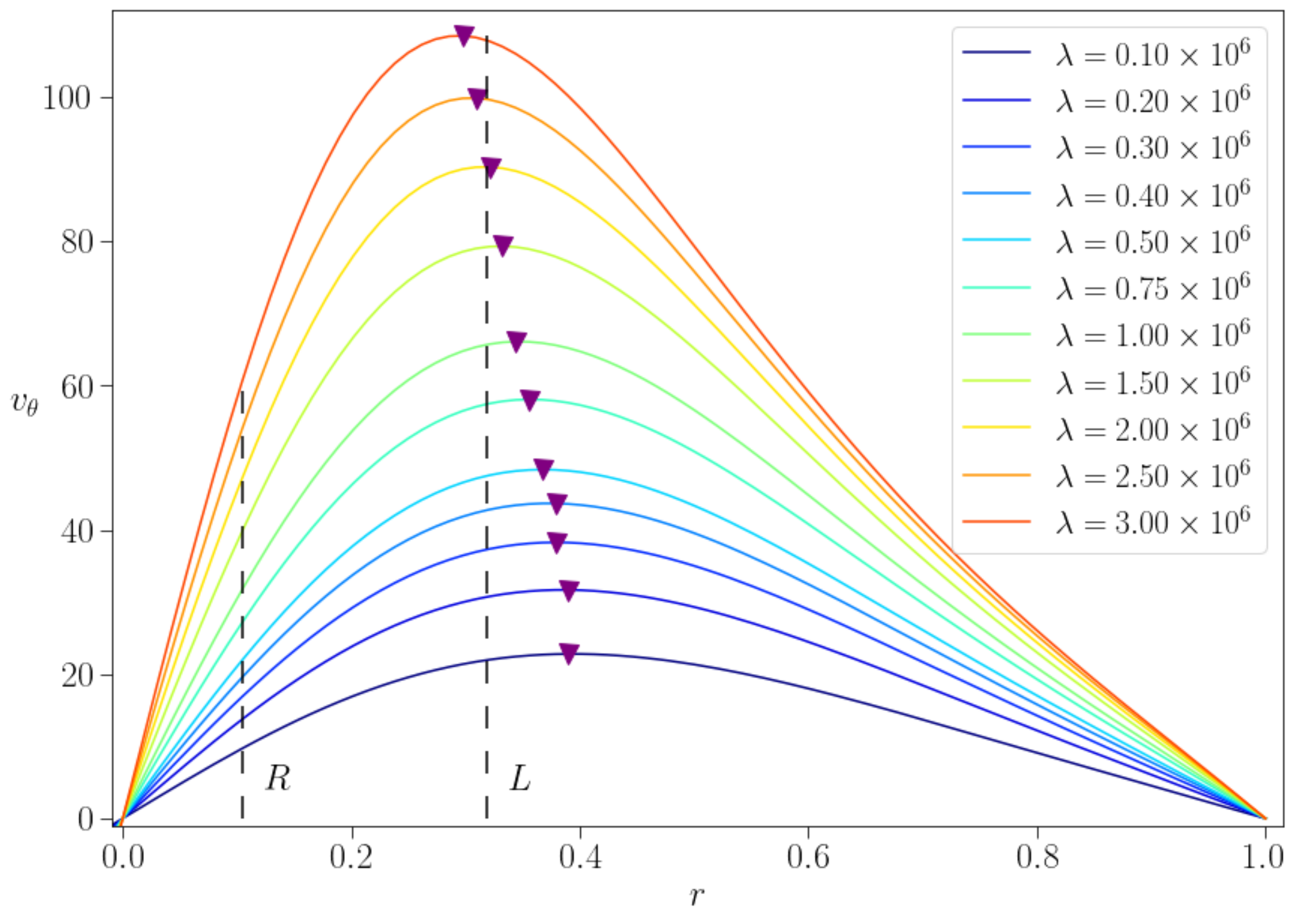}
\caption{The instanton profiles of $v_\theta(r,z=0)$ at the final evolution time for several values of $\lambda$, and 
$R/L\approx0.32$. The vertical dashed lines give the radial positions $r=R$ and $r=L$, while the inverted
triangles locate the peak positions of $v_\theta(r,z=0)$.}\label{fig5}
\end{figure}
\vspace{0.2cm}

It is not difficult to see, from Fig.~\ref{fig4}, that the azimuthal component of the velocity field is not peaked at $r=R$ as we have found in our analysis of the filtered DNS instantons (extreme circulation events), previously summarized in Fig.~\ref{fig3}. This is not a completely casual remark. In fact, it turns out that the coupled instanton equations (\ref{instv}-\ref{instcirc}) become very stiff at large values of $\lambda$, precisely where we expect to model asymptotically large circulation fluctuations.

In Fig.~\ref{fig5} we show that, as $\lambda$ grows, the peak position of $v_\theta(r,z=0)$ slowly drifts to the left (smaller values of $r$). A rough estimate indicates that a peak at $r=R$ would be reached for $\lambda \sim 10^{10}$, which is several orders of magnitude beyond the domain of convergence achieved in our applications of the Chernykh-Stepanov method.

We find, thus, that direct numerical schemes to solve the instanton equations (\ref{instv}-\ref{instcirc}), like the ones we have applied, are able to address the parabolic cores of cPDFs, while the description of their tails remains challenging. In order to circumvent the stiffness of the instanton equations, and to model the far cPDF tails, space-time reparametrizations \cite{chernykh2001large,grafke2014arclength,grigorio2020parametric} and further large deviation techniques are likely to be necessary, as illustrated by the use of hybrid Monte Carlo algorithms, so far only applied to Burgers turbulence \cite{margazoglou2019hybrid}.

%The connection between the statistics of velocity circulation and vorticity is clear for very small scales, where the vorticity field is smooth and circulations is simply the product of vorticity by the area of the contour, in this limit we claim our numerical solution to be closely related to the vorticity instanton. For this reason we performed numerical experiments, calibrating $\lambda$ to find $\omega_z(r=0,z=0)\approx60$ to directly compare vorticity field profile to the one found in found in \cite{schorlepp2021spontaneous}. Even though the governing equations are different for those two observable, Fig.~(\ref{fig7}) shows clear similarities between the solutions mainly on $\omega_z$ and $\omega_r$. A considerable decreasing on the intensity of $\omega_\theta$ is seen when comparing the two solutions (about ten times lower) and a third ring structure appears close to the boundary. We argue that although very small, the contour radius cannot be completely regarded as asymptotic and finite contour effects can take a place on the solution. 
%\begin{figure}[H]
%\centering\includegraphics[width=0.9\linewidth]{comparison.pdf}
%\caption{Vorticity contour plot for $N=128$, $M=256$, $N_t=400$, $dt=1.25\ 10^{-3}$, $R\approx0.02 L$ and $\lambda=666000$.}\label{fig7}
%\end{figure}

\section{Conclusions}\label{sec4}

We have investigated the occurrence of extreme circulation events in three-dimensional homogeneous and isotropic turbulence. We show that the time evolution of circulation, conditionally averaged to the observation of prescribed large deviation events can be effectively modeled with the help of eddy-viscosity phenomenology combined with instanton functional techniques. Our modeling analysis is corroborated from a careful treatment of large DNS databases, at various Reynolds numbers.

Proceeding with the same set of extreme circulation events, we have furthermore inspected time-dependent conditionally averaged velocity field configurations. Resulting axisymmetric vortex structures clearly arise from these filtering procedures. The comparison with the eddy-viscous (creeping) instantons is only reasonable at the core of the flow configurations (circulation instantons), which, nevertheless, is all one needs to get accurate averages of the time-dependent instanton circulations ending in extreme events.

An extensive numerical effort to solve the instanton equations suggests that extreme circulation events have an interesting underlying topological structure. Axisymmetric circulation instantons are composed of a main vortex, which is at the center of a surrounding pair of counter-rotating vortex rings. The existence of this triple vortex structure drives our attention to possible formal connections with superfluidity. It is known that the motion of a superfluid vortex ring is followed by a companion pair of normal vortex rings \cite{kivotides2000triple}. We note, then, that the instanton Eqs.~(\ref{instv}) and (\ref{instp}) are actually analogous to the HVBK hydrodynamic equations describing the self-induced propagation of a quantum vortex ring.

Although the employed numerical databases allowed us to validate a viscous-eddy modeling of extreme circulation events, they are not large enough to disclose the topological features of the circulation instantons. Larger databases are required, in order to accomplish this challenging task. Another related important issue, deserved for further studies, has to do with the implementation of algorithmic improvements in the numerical solution of the instanton equations, in order to model the far cPDF tails \cite{iyer2019circulation,iyer2021area} and the off-core azimuthal velocity component of the circulation instantons. It is worth emphasizing that some alternatives have been already applied with success to the paradigmatic example of Burgers turbulence \cite{grafke2014arclength,grafke2015relevance}.

Intense vortex structure is known to have an important role in turbulence at high Reynolds numbers. Such structures can be modeled by an interacting molecular gas when statistical properties of circulation in 2D slices of turbulent flows are analyzed \cite{apolinario2020vortex,moriconi2021multifractality,moriconi2022statistics,pereira2022hard}. A possible source of future investigation is to explore a single or a bunch of modeled structures in order to compare its spatial distribution to the instanton solutions.

\section*{Acknowledgement}

This work has been partially supported by CAPES via Grant No. 88887.336246/2019-00 (V.J.V.) and by the Simons Foundation Award ID 651475 (G.B.A.).

% \bibliographystyle{model1-num-names}
% \bibliography{biblio.bib}

\appendix*

\section{Spectral Approach to the Instanton Equations in Cylindrical Coordinates}

In order to numerically solve the set of Eqs.~(\ref{instv}-\ref{instcirc}) we rewrite them in cylindrical coordinates. In light of the nonlinear nature of hydrodynamic interactions, there would be no a priori reason to consider axisymmetric solutions in detriment of any other. Nevertheless, we expect axisymmetric solutions to dominate extreme events since the forcing term is invariant under rotations around the $z$-axis. Taking this symmetry into account, Eqs.~(\ref{instv}-\ref{instcirc}) read
\be\label{axisinst}
\begin{cases}
    \hat{\mathcal{L}}^-v_r+v_r/r^2-v_\theta^2/r+\partial_rP=(\tilde{\chi}\star p_r)\ , \\
    \hat{\mathcal{L}}^-v_\theta+v_\theta/r^2+v_\theta v_r/r=(\tilde{\chi}\star p_\theta)\ , \\
    \hat{\mathcal{L}}^-v_z+\partial_zP=(\tilde{\chi}\star p_z)\ , \\
    \partial_r(rv_r)/r+\partial_zv_z=0\ , \\
    \hat{\mathcal{L}}^+p_r-p_r/r^2-v_\theta p_\theta/r+v_r\partial_r p_r+v_\theta\partial_r p_\theta+v_z\partial_r p_z+\partial_rQ=0\ , \\
    \hat{\mathcal{L}}^+p_\theta-p_\theta/r^2+(2v_\theta p_r-v_rp_\theta)/r=\lambda\delta(t)\delta(r-R)\delta(z)\ , \\
    \hat{\mathcal{L}}^+p_z+v_r\partial_z p_r+v_\theta\partial_z p_\theta+v_z\partial_z p_z+\partial_zQ=0\ , \\
    \partial_r(rp_r)/r+\partial_zp_z=0\ , \ 
\end{cases}
\ee
where $\hat{\mathcal{L}}^\pm\psi=\partial_t\psi+v_z\partial_z\psi+v_r\partial_r\psi\pm(\partial_z^2\psi+\partial_r(r\partial_r\psi)/r)$ and $(\phi\star \psi)$ is the spatial convolution between $\phi$ and $\psi$. We implement a pseudo-spectral approach where the fields are expanded in a truncated Fourier-Chebyshev series of orders $N$ and $M$, respectively. The collocation points are defined as
\be
z_j=\frac{2\pi j}{N} \ , \ 
\ee
and
\be
r_m=\cos{\left(\frac{\pi m}{M-1}\right)}\ , \ 
\ee
where, $j$ and $m$ are integers in the ranges $j \in [-N/2,N/2-1]$ and $m \in [0,M-1]$. Along the longitudinal direction, usual collocation points for Fourier series are used, while for the radial direction we employ Gauss-Lobatto collocation points, corresponding to the extrema of the $M^{th}$ order Chebyshev polynomials \cite{canuto1988spectral}.
This choice of a non-homogeneous grid hinders a direct comparison among simulations with different resolutions. However, it allows one to adopt a Fast Fourier Transform (FFT) algorithm to compute nonlinear terms, which are otherwise computationally expensive \cite{peyret2013spectral}. Moreover, for even $M$ the coordinate singularity at $r=0$ is explicitly avoided on the collocation, with the drawback of requiring the radial inverval to be duplicated, as $r \in [0,1]\rightarrow r\in [-1,1]$, as a requirement of the Chebyshev polynomial expansion.

All fieds in Eq.~\refp{axisinst} are given by a truncated polynomial series of order $N$ in the Fourier basis and $M$ in the Chebyshev basis,
\be
\phi(r_m,z_j)\equiv\phi_{mj}=\sum_{l=0}^{M-1}\sum_{k=-N/2}^{N/2-1}\hat{\phi}_{lk}T_l(r_m)e^{ikz_j}\ , \ 
\ee
where $T_l(x)$ is the $l^{th}$ Chebyshev polynomial. Exceptions are the pressure fields $Q(\vec{x},t)$ and $P(\vec{x},t)$, which are of order $M-2$ in the Chebyshev expansion. This guarantees the stability of the numerical method, an approach known as the $\mathbb{P}_N-\mathbb{P}_{N-2}$ approximation \cite{peyret2013spectral}, further discussed ahead. Radial derivatives of these fields are readily computed using recursion relations of the Chebyshev polynomials, namely,
\be
(\partial_r\phi)_{mj}=\sum_{l=0}^{M-1}D_{ml}\phi_{lj}\ , \ 
\ee
where the $M\times M$ matrix $D$ has, for all fields but the pressure related ones, the entries
\be\label{Dv}
D_{mj}=
\begin{cases}
    (-1)^{m+j}c_m/(c_j(r_m-r_j)) &\ , \  m\neq j\\
    -r_m/(2(1-r_m^2)) &\ , \  m = j, \ m,j\neq [0,(M-1)] \\
    \pm (2(N-1)^2+1)/6 &\ , \  m=j=0 \text{ or } m=j=(M-1),
\end{cases}
\ee
with $c_0=c_{M-1}=2$ and $c_i=1$ otherwise. As for the pressure fields, which are 2 degrees lower in the Chebyshev expansion, they must be interpolated from the $M-2$ Gauss-Lobatto grid to the $M$ grid where the other fields are defined. This can be done through a Lagrange interpolation (see Ref.~\cite{peyret2013spectral} for details), giving
\be\label{D2p}
D^{PQ}_{mj}=
\begin{cases}
    (-1)^{j+m}(1-r_j^2)/((r_m-r_j)(1-r_m^2)) &\ , \  m\neq j, \ m,j \in [1,M-2] \\
    3 r_m/(2(1-r_m^2)) &\ , \ m=j, \ m \in [1,M-2].
\end{cases}
\ee

% The above $M\times M$ matrix $D$ is however different for pressure fields, since they are 2 degrees lower in the Chebyshev expansion, they must be interpolated from the $M'=M-2$ Gauss-Lobatto grid to the $M$ grid where the other fields are defined. This can be done through a Lagrange interpolation which maps the derivative matrix in the grid $r'_j$ with $j\in[0,M'-1]$ into its values in the grid $r_j$ with $j\in[1,M-2]$ (for details see Ref. \cite{peyret2013spectral}). The explicit form of these matrices are the following,
% \be\label{Dv}
% D_{mj}=
% \begin{cases}
%     (-1)^{m+j}c_m/(c_j(r_m-r_j)) &\ , \  m\neq j\\
%     -r_m/(2(1-r_m^2)) &\ , \  m = j, \ m,j\neq [0,(M-1)] \\
%     \pm (2(N-1)^2+1)/6 &\ , \  m=j=0 \text{ or } m=j=(M-1)
% \end{cases}
% \ee
% \be\label{D2p}
% D^{PQ}_{mj}=
% \begin{cases}
%     (-1)^{j+m}(1-r_j^2)/((r_m-r_j)(1-r_m^2)) &\ , \  m\neq j, \ m,j \in [1,M-2] \\
%     3 r_m/(2(1-r_m^2)) &\ , \ m=j, \ m \in [1,M-2]
% \end{cases}
% \ee
% where $c_i=2$ if $i=0$ and $i=M-1$ or $c_i=1$ otherwise. 

By virtue of the explicit form of these matrices, $r$-derivatives are evaluated in physical space while $z$-derivatives are better performed in Fourier space. Ultimately, in order to have consistently represent all fields in the mirrored radial domain, they must satisfy the following reflection properties:
\begin{equation}
\begin{cases}
    &\phi_{mj}=\phi_{(-m)j} \text{ for scalar fields,} \\
    &\phi_{z,mj}=\phi_{z,(-m)j} \text{ for the }z\text{-component of vector fields} \\
    &\phi_{\beta,mj}=-\phi_{\beta,(-m)j} \text{ for the radial and azimuthal components (}\beta \text{ is either } r\text{ or }\theta).
\end{cases}
\end{equation}

The time discretization follows a combined Adams-Bashforth/Implicit Backward Differentiation method of second order (AB/BDI2) \cite{peyret2013spectral}, which consists on implicit evaluations of the linear terms and explicit evaluations of the nonlinear terms.
As an illustration, consider the equation
\be
\partial_t \phi=L(\phi)+N(\phi)+J\ , \
\ee
where, $L(\phi)$ and $N(\phi)$ stand for the linear and nonlinear terms of the differential equation, respectively, and $J$ accounts for a forcing term or pressure gradient. The above equation is discretized at regularly spaced time instants $t_n=n dt$ as
\be\label{examp}
\frac{3\phi^{(n+1)}-4\phi^{(n)}+\phi^{(n-1)}}{2dt}
=L\phi^{(n+1)}
+2N(\phi^{(n)})-N(\phi^{(n-1)})
+J^{(n+1)}\ , \ 
\ee
where $\phi^{(n)}=\phi(t_n)$. In the first time step, we set $\phi^{(-1)}=\phi^{(0)}$ and change $dt\rightarrow 3dt/2$, reducing this step to the usual Euler scheme.
The choice of discretizing the pressure gradient terms as $\partial_i P^{(n+1)}$ leads to a Stokes problem which is solvable by the Uzawa method \cite{uzawa1958iterative}. We note that similar discretization setups were successfully applied to the Navier-Stokes equations in cylindrical coordinates with a few different boundary conditions \cite{raspo2002spectral,lopez1998efficient,peres20123d}. We also remark that the nonlinear terms are properly de-aliased following a standard 3/2-rule.

Applying the above discretization procedure to the system of Eqs.~\refp{axisinst} supplemented by Dirichlet boundary conditions, one finds a system of coupled equations for each independent Fourier mode $k\in[N/2,N/2-1]$ for the forward and backward time integration of the velocity and conjugate field, respectively,
\be\label{discinst}
\begin{cases}
    \hat{L}_{r,k}\ket{v_{r,k}^n}=-\hat{D}^{PQ}\ket{P_k^n}+\ket{f_{r,k}^n} &\ , \ j\in [1,M-2]\ , \\
    \hat{L}_{r,k}\ket{v_{\theta,k}^n}=\ket{f_{\theta,k}^n} &\ , \ j\in [1,M-2]\ , \\
    \hat{L}_{z,k}\ket{v_{z,k}^n}=-ik\ket{P_k^n}+\ket{f_{z,k}^n} &\ , \ j\in [1,M-2]\ , \\
    \hat{R}^{-1}\hat{D}\hat{R}\ket{v_{r,k}^n}+ik\ket{v_{z,k}^n}=0 &\ , \ j\in [0,M-1]\ , \\
    (\ket{v_{r,k}^n},\ket{v_{\theta,k}^n},\ket{v_{z,k}^n})=0 &\ , \  j=0\text{ and }j=M-1\ , \\
    \\
    \hat{L}_{r,k}\ket{p_{r,k}^n}=\hat{D}^{PQ}\ket{Q_k^n}+\ket{g_{r,k}^n} &\ , \ j\in [1,M-2]\ , \\
    \hat{L}_{r,k}\ket{p_{\theta,k}^n}=\ket{g_{\theta,k}^n} &\ , \ j\in [1,M-2]\ , \\
    \hat{L}_{z,k}\ket{p_{z,k}^n}=ik\ket{Q_k^n}+\ket{g_{z,k}^n} &\ , \ j\in [1,M-2]\ , \\
    \hat{R}^{-1}\hat{D}\hat{R}\ket{p_{r,k}^n}+ik\ket{p_{z,k}^n}=0 &\ , \ j\in [0,M-1]\ , \\
    (\ket{p_{r,k}^n},\ket{p_{\theta,k}^n},\ket{p_{z,k}^n})=0 &\ , \ j=0\text{ and }j=M-1\ , \\
\end{cases}
\ee
where the ket notation represents the $N$-dimensional vector $\ket{\phi_{i,k}^n}=[\phi_{i,jk}^n]$ with $j\in [0,M-1]$. The $M\times M$ matrices $\hat{R}=\mathrm{diag}(r_0,r_1,\cdots,r_{M-1})$, $\hat{L}_{r,k}=(3/(2|dt|)+k^2)\hat{\mathbb{I}}-\hat{D}^2-\hat{R}^{-1}\hat{D}+\hat{R}^{-2}$ and $\hat{L}_{z,k}=\hat{L}_{r,k}-\hat{R}^{-2}$ can be efficiently inverted and stored in a pre-processing stage. The $M$ vectors $(\ket{f_{r,k}^n},\ket{f_{\theta,k}^n},\ket{f_{z,k}^n})$ and $(\ket{g_{r,k}^n},\ket{g_{\theta,k}^n},\ket{g_{z,k}^n})$ are, respectively, the explicit part of the discretized Eqs.~\ref{instv} and \ref{instp}. For instance, in Eq.~\ref{examp} one has $\ket{f_k^n}=(4\phi^{(n-1)}-\phi^{(n-2)}+2N(\phi^{(n-1)})-N(\phi^{(n-2)}))/3$.

The algebraic system defined by Eq.~\ref{discinst} has a unique solution for every $k\neq0$. The $\mathbb{P}_N-\mathbb{P}_{N-2}$ approximation prevents zero eigenvalues of the Uzawa operator\footnote{The Uzawa operator is obtained by solving formally the pressure field by setting the momentum equations into the incompressibility constraint, in this case $\hat{Z}_k=\hat{R}^{-1}\hat{D}\hat{R}\hat{L}_r^{-1}\hat{D}^{PQ}-k^2\hat{L}_z^{-1}$.} for $k=0$ and avoids the requirement of prescribing boundary conditions to the pressure field. Indeed, non-uniqueness of the solution is related to the fact that pressure fields are defined up to a constant and a simple calculation shows that the unique solution consistent with the boundary conditions and the incompressibility constraint for $k=0$ is $\ket{v_{r,0}^{n}}=0$.

The convolutions in Eq.~\eqref{axisinst} and in the action integral (Eq.~\ref{actionre-scaled})
can be efficiently computed using the explicit form of the Fourier transformed correlation $\tilde{\chi}$ and the inverted Chebyshev derivative matrix, 
\be\label{conv}
\mathrm{FFT}_z\big[(\tilde{\chi}\star p_{\beta})\big]_{jk}^{(n)}=(2\pi)^{3/2} L e^{-\frac{r_j^2}{2L^2}}
e^{-\frac{k^2L^2}{2}}\sum_{l}\left(\hat{D}^{-1}_{M/2,l}-\hat{D}^{-1}_{0,l}\right)U_{jl,\beta}
f^{(n)}_{l,\beta}\ , \
\ee
\be\label{intact}
S=(2\pi)^{7/2} L \sum_{n,k,l,j,\beta}e^{-\frac{k^2 L^2}{2}}
\left(\hat{D}^{-1}_{M/2,l}-\hat{D}^{-1}_{0,l}\right)\left(\hat{D}^{-1}_{M/2,j}-\hat{D}^{-1}_{0,j}\right)
f^{(n)}_{lk,\beta}
U_{jl,\beta}
\left(f^{(n)}_{jk,\beta}\right)^{\star}\ , \
\ee
where $\beta=(r,\theta,z)$, $f^{(n)}_{lk,\beta}=r_l \exp{(-r_l^2/2L^2)}FFT_z[p_\beta]^{(n)}_{lk}$, $U_{jk,\beta}=I_0(r_jr_k/L^2)$ for $\beta=z$, and $U_{jk,\beta}=I_1(r_jr_k/L^2)$ for $\beta=(r,\theta)$, with $I_0,I_1$ being modified Bessel functions of the first kind. 

\begin{figure}[h]
\centering\includegraphics[width=0.9\linewidth]{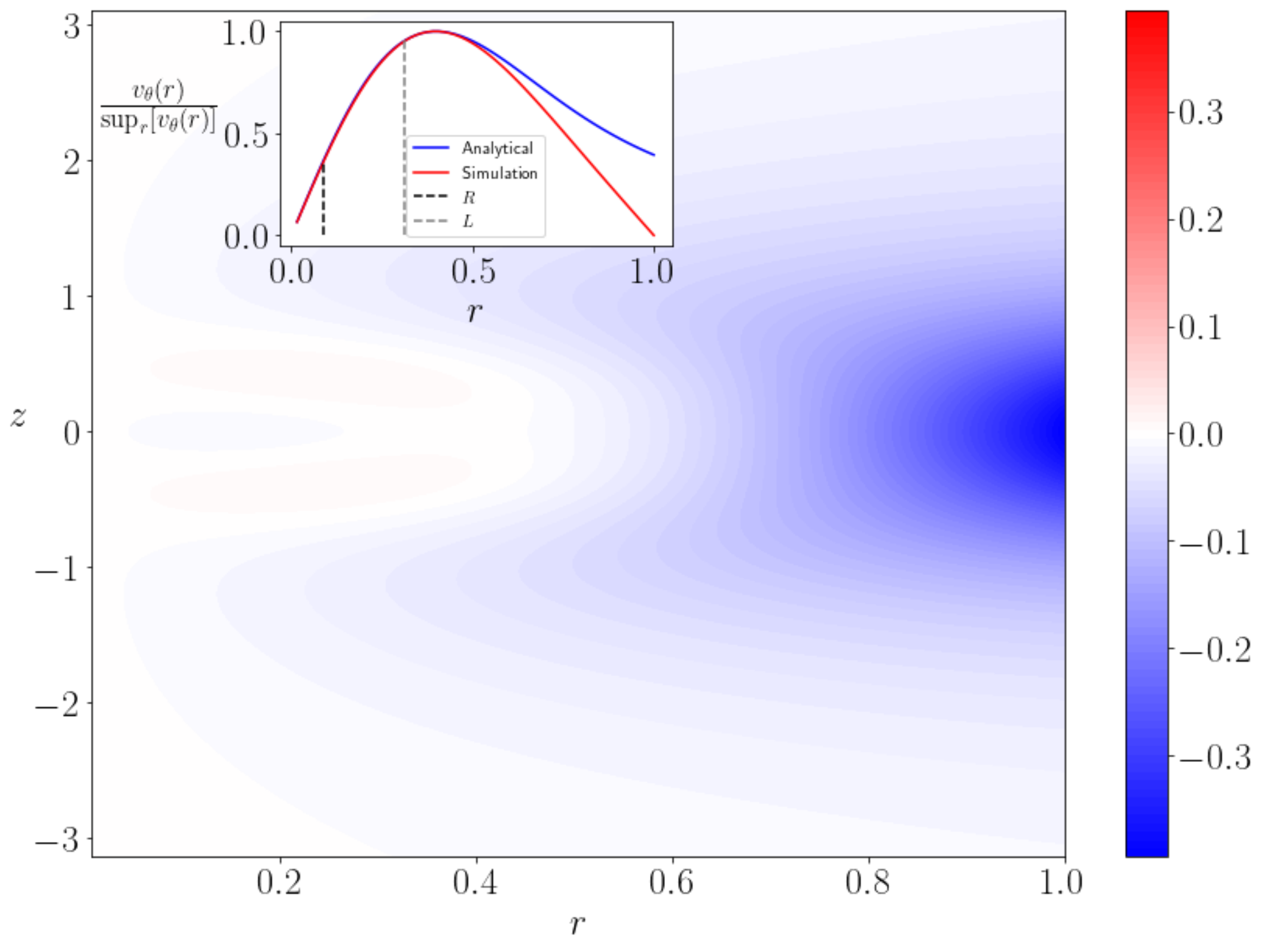}
\caption{Difference between normalized numerical and analytical solutions of the azimuthal velocity component at $t=0$ as functions of $r$ and $z$, for $R\approx0.09$ and $L=1/\pi$. Inset: profiles for $z=0$.}\label{fig6}
\end{figure}

In order to validate the numerical method, we performed several numerical experiments of the linear instanton. Fig.~\ref{fig6} compares the $v_\theta$ component of the numerical solution with the analytical result from Eq.~\ref{velcomp}. We first note that the Dirichlet boundary condition has a relevant influence on the solution for $r \gtrsim 0.5$. This is not a surprising though, since Eq.~\ref{velcomp} holds for unbounded domains, thus a slower decay is expected. Variations of about $\pm 5\%$ in the peak position are also observed depending on how distant to the boundaries one sets $L$ and $R$. We fixed $L=1/\pi$ to minimize such boundary effects.

The viscous action was calculated by means of Eq.~\ref{intact}, and it was found to be compatible with $S=\Gamma^{\alpha_\Gamma}/(2\sigma^2)$ with $\alpha_\Gamma=1.99999993(5)$ and $\sigma=(\beta_\sigma/N) R^{1.95(2)}$, with errors estimated by averaging results obtained with different grid resolutions, both in time and space. Small variations of $\beta_\sigma$ are seen when $dt$ and/or $M$ are changed, but one must keep in mind that the radial collocation is not regular, and hence direct comparisons among different resolutions in $M$ are not perfect. As for the $dt$ dependence, finite time effects are in play, since decreasing $dt$ for a fixed number of timesteps $N_t$ also causes the total simulation time to decrease, so the boundary conditions $p_i(\vec{x},T)=v_i(\vec{x},T)=0$ are effectively imposed on different time instants $T=-N_t dt$. In out tests, we worked with all combinations of $N_t=100$, $200$, and $400$, with $dt=0.0025$, $0.005$, and $0.01$.

The conclusion drawn from this set of numerical experiments is that both spatial and statistical properties are accurately captured by the numerical algorithm used to solve the instanton equations, at least in the linear approximation.

\end{document}